\documentclass[12pt,a4paper]{article}

  \usepackage{a4wide}
  \usepackage{latexsym}
  \usepackage{epsf}
  \usepackage{amssymb}
  \usepackage{graphicx}
  \usepackage{amsmath, cite}
  \usepackage{bbm}
  \usepackage{bbm}
  \usepackage{amsmath,amssymb,amsthm}
  \usepackage{pst-all}

\renewcommand{\d}{\textrm{d}}
\newcommand{\Real}{\textrm{I\!R}}

\newcommand{\e}{\textrm{e}}
\renewcommand{\d}{\textrm{d}}

\newcommand{\ISO}{\mathop{\rm ISO}}

\newcommand{\rf}[1]{(\ref{#1})}

\begin{document}

\begin{flushright}
\small
KUL-TF-07-28\\
UGFT-221-07 \\
CAFPE-91-07
\date \\
\normalsize
\end{flushright}
\vspace{0.5cm}
\begin{center}

{\LARGE \bf{A first-order formalism for timelike\\\vspace{0.8cm}
 and spacelike brane solutions}}\\

\vspace{1.3cm} {\large Bert Janssen$^{\dagger}$, Paul
Smyth$^{\ast}$,
Thomas Van Riet$^{\ddagger}$ and Bert Vercnocke$^{\ddagger}$} \\[3mm]

\vspace{1cm}

$\dagger$\textit{ Departamento de F\'isica Te\'orica y del Cosmos\\
and Centro Andaluz de F\'isica de Part\'iculas Elementales\\
Universidad de Granada, 18071 Granada, Spain}\\
{\small\upshape\ttfamily bjanssen@ugr.es}\\[3mm]
\textit{$^\ast$ II. Institut f\"ur theoretische Physik der
Universit\"at Hamburg\\
Luruper Chaussee 149, 22761 Hamburg, Germany} \\
{\small\upshape\ttfamily Paul.Smyth@desy.de }\\[3mm]
$\ddagger$\textit{ Institute for Theoretical Physics, K.U. Leuven,\\
Celestijnenlaan 200D, B-3001 Leuven, Belgium}
\\{\small\upshape\ttfamily
Bert.Vercnocke,Thomas.VanRiet@fys.kuleuven.be }\\[3mm]

\vspace{2cm}

{\bf Abstract}
\end{center}

\begin{quotation}

\small We show that the construction of BPS-type equations for
non-extremal black holes due to Miller et.~al.~can be extended to
branes of arbitrary dimension and, more importantly, to
time-dependent solutions. We call these first-order equations
\textit{fake}- or \textit{pseudo}-BPS equations in light of the
formalism that has been developed for domain wall and cosmological
solutions of gravity coupled to scalar fields. We present the
fake/pseudo-BPS equations for all stationary branes (timelike
branes) and all time-dependent branes (spacelike branes) of an
Einstein-dilaton-$p$-form system in arbitrary dimensions.
 \end{quotation}

\newpage

\pagestyle{plain}
\tableofcontents
\section{Introduction}

Many brane-type solutions have been constructed for simple
truncations of supergravity theories to the form (see, for
instance, \cite{Stelle:1998xg})
\begin{equation}\label{bosonicaction}
S=\int \d^D x\sqrt{-g}\Bigl( \tfrac{1}{2\kappa_D^2}\mathcal{R}-
\tfrac{1}{2}(\partial\phi)^2-\tfrac{1}{2 n!}\e^{a\phi}F_n^2
\Bigr)\,,
\end{equation}
where $\mathcal{R}$ is the Ricci-scalar, $\kappa_D^2$ is the
$D$-dimensional gravitational coupling, $\phi$ is a scalar field (the dilaton)
and $F_n$ is the field strength of some $(n-1)$-form, $F_n=\d
A_{n-1}$ if $n>0$. For the special case $n=0$ we consider $F_n^2$ to
be a cosmological term (scalar potential). The parameter $a$ is
fixed and is called the dilaton coupling.

The equations of motion derived from this action admit electrically
charged ($n-2$)-branes and magnetically charged ($D-n-2$)-branes. A
brane solution can be stationary or time-dependent. The metric of a
stationary $p$-brane is given by
\begin{equation}\label{stationair}
\d s^2_D=\e^{2A(r)}\eta_{\mu\nu}\d x^{\mu}\d x^{\nu} + \e^{2B(r)}\d
r^2 + \e^{2C(r)}\d\Sigma_k^2\,,
\end{equation}
where $\eta$ is the usual Minkowski metric in $p+1$ dimensions,
$\eta=\text{diag}(-,+, \ldots, +)$ and $\d\Sigma_k^2$ is the metric
of a $d$-dimensional maximally symmetric space with unit curvature
$k=-1,0,1$, such that the Ricci scalar is given by ${\cal R}_d = k d
(d-1)$.
 When $k=1$ the solutions possess a rotational
symmetry and can be asymptotically flat (in contrast to $k=-1$). For
$D=10$ and specific values of $a$ and $n$ the solutions correspond
to D-branes in string theory.

The metric of the time-dependent branes is similar
\begin{equation}\label{timedependent}
\d s^2_D=\e^{2A(t)}\delta_{\mu\nu}\d x^{\mu}\d x^{\nu} -\e^{2B(t)}\d
t^2 + \e^{2C(r)}\d\Sigma^2_k\,,
\end{equation}
where $\delta_{\mu\nu}$ is the usual flat Euclidean metric in $p+1$
dimensions, $\delta_{\mu\nu}=\text{diag}(+,+ \ldots +)$. In the
$k=-1$ case the transverse space possesses a Lorentzian symmetry and
can be asymptotically flat (in contrast to $k=+1$ solutions). These
solutions are the spacelike branes (S-branes) introduced by Gutperle
and Strominger \cite{Gutperle:2002ai}, who conjectured that such
branes correspond to specific time-dependent processes in string
theory.

From now on we shall call the stationary branes with spherical
slicing ($k=+1$) \emph{timelike} branes and the time-dependent
branes with hyperbolic slicing ($k=-1$) \emph{spacelike} branes.
All the other possible slicings are also covered
here, but we choose to highlight only these two cases.

It has been known for a long time that particular timelike $p$-brane
solutions of supergravity preserve some fraction of supersymmetry.
Practically this means that the solutions fulfill some
\emph{first-order} differential equations that arise from demanding
the supersymmetry transformations to be consistently satisfied for
vanishing fermions. Such first-order equations have become known as
Bogomol'nyi or BPS equations, after Bogomol'nyi's
\cite{Bogomolny:1975de}, and Prasad and Sommerfield's
\cite{Prasad:1975kr} work on first-order equations and exact
solutions for magnetic monopoles in the Yang--Mills--Higgs theory.
It was then later shown that this limit is intimately linked to the
preserved supersymmetry of solitons in supersymmetric theories by
Witten and Olive \cite{Witten:1978mh}. The term \textit{BPS
equation} is now generically used for equations of motion that are
inferred by rewriting the action as a sum of squares. Supersymmetric
solutions, in general, belong to this class. Stationary non-extremal
and time-dependent solutions cannot preserve supersymmetry in
ordinary supergravity theories. Naively one therefore expects that
such solutions cannot be found from BPS equations, but rather by
solving the full second-order equations of motion.

To our knowledge there are three instances in the literature
where it has been shown that this view is too pessimistic:\\

(I) Not all extremal black hole solutions of supergravity have to
be supersymmetric. It turns out that many non-supersymmetric but
extremal solutions fulfill first-order equations in a given
supergravity theory (see for instance \cite{Ceresole:2007wx,
Andrianopoli:2007gt, LopesCardoso:2007ky}). More surprisingly,
Miller et.~al have shown that the \emph{non-extremal}
Reissner--Nordstr\"om black hole solution of Einstein--Maxwell
theory can be found from first-order equations \cite{Miller:2006ay},
by a clever rewriting of the action as a sum of squares \`a la
Bogomol'nyi. The method of \cite{Miller:2006ay} is the main tool for
the present paper.

(II) Many stationary domain wall solutions that do not preserve any
supersymmetry have been shown to allow for first order-equations by
the construction of a \emph{fake} superpotential
\cite{Freedman:2003ax, Celi:2004st, Papadimitriou:2006dr}. The
domain walls in question are solutions to the following Lagrangian
\begin{equation}\label{action}
\mathcal{L}=\sqrt{-g}\Bigl(\mathcal{R} -
\tfrac{1}{2}G_{ij}(\Phi)g^{\mu\nu}
\partial_{\mu}\Phi^i\partial_{\nu}\Phi^j - V(\Phi)\Bigr)
\end{equation}
where $\Phi^i$ are scalars fields, $G_{ij}(\Phi)$ the metric on the
target space and $V(\Phi)$ the scalar potential. The metric Ansatz
for a flat domain wall is\footnote{For
simplicity we only discuss flat cosmologies and flat domain walls.}
\begin{equation}
\d s^2= \e^{2B(z)}\d z^2+ \e^{2A(z)}\eta_{ab}\d x^a\d x^b\,,
\end{equation}
where $\eta_{ab}$ is diag$(-,+,\ldots,+)$. The high degree of
symmetry of this Ansatz is only consistent when the fields that
support the solution depend solely on the $z$-coordinate i.e.
$\Phi^i=\Phi^i(z)$. We then suppose that a scalar function $W(\Phi)$
exists such that
\begin{equation}\label{susy}
V=\tfrac{1}{2}G^{ij}\partial_i W\partial_j W -
\tfrac{D-1}{4(D-2)}W^2 \,,
\end{equation}
which allows the action to be written as a sum of squares
(neglecting boundary terms) \cite{Skenderis:1999mm}
\begin{equation} S=\int\d z\,
\e^{(D-1)A+B}\Bigl\{\tfrac{(D-1)}{4(D-2)}\bigl[W-2(D-2)\e^{-B}A^\prime
\bigr]^2
-\tfrac{1}{2} \Bigl| \! \Bigl|  \e^{-B} (\Phi^i)^\prime + G^{ij}\partial_j W \Bigr| \! \Bigr|^2 \Bigr\}\,,\\
\end{equation}
where a prime denotes a derivative with respect to $z$. Solutions
are obtained when each square in the action is zero. If $W$ is a
superpotential of some supersymmetric theory, these first-order
equations are the standard BPS equations for domain walls that would
arise by demanding that the supersymmetry variations are satisfied
for vanishing fermions. However, for \textit{every} $W$ that obeys
(\ref{susy}) we can find a corresponding DW-solution. If $W$ is not
related to the quantity appearing in the supersymmetry
transformations the resulting solutions are called \textit{fake
supersymmetric}.

(III) FLRW-cosmologies are very similar to domain walls
\cite{Cvetic:1994ya, Skenderis:2006fb, Skenderis:2006jq}, the
difference in metrics being given by a few signs
\begin{equation}
\d s^2= -\e^{2B(t)}\d t^2 + \e^{2A(t)}\delta_{ab}\d x^a\d x^b\,.
\end{equation}
When the relation (\ref{susy}) is changed by an overall minus-sign
\begin{equation}
V=-\tfrac{1}{2}G^{ij}\partial_i W\partial_j W +
\tfrac{D-1}{4(D-2)}W^2 \,,
\end{equation}
the same first-order equations for domain walls exist for
cosmologies, where now the primes indicate derivatives with respect
to time. These relations have become known as \emph{pseudo-BPS}
conditions \cite{Skenderis:2006fb,Skenderis:2006jq} (see also
\cite{Salopek:1990jq, Bazeia:2005tj} for the first-order framework
in cosmology). As for domain walls one readily checks that these
first-order equations arise from the fact that the action can be
written as a sum of squares \cite{Chemissany:2007fg}. The structure
underlying the existence of these first-order equations can be
understood from Hamilton--Jacobi theory \cite{Skenderis:2006rr,
Townsend:2007aw, Townsend:2007nm }.\footnote{In ordinary
supergravity theories the pseudo BPS relations cannot be related to
supersymmetry preservation. However, in the case of supergravity
theories with `wrong sign' kinetic terms the pseudo-BPS relations
are related to true supersymmetry \cite{deWit:1987sn,
Behrndt:2003cx, Bergshoeff:2007cg, Skenderis:2007sm,Vaula:2007jk}. In this paper
we shall consider ordinary supergravity theories and therefore
pseudo-BPS conditions are not related to supersymmetry. Practically
this means that we have first-order equations which can be
understood to originate from a Bogomol'nyi rewriting of the
action.} \\

The examples given above (I-III) are only a subset of the different
$p$-branes that exist, namely timelike $0$-branes in $D=4$ (the RN
black holes) and $(D-2)$-branes (domain walls and FLRW-cosmologies).
It is the aim of this paper to understand in general when stationary
and time-dependent $p$-branes in arbitrary dimensions can be found
from BPS equations.

One of the main subtleties that arises in this generalisation is
that there exist two kinds of black deformations of timelike
$p$-branes which coincide for the special case of black holes in
four dimensions. For this reason the example treated by Miller
et.~al.~\cite{Miller:2006ay} is not completely representative.
Secondly, for time-dependent solutions, it has yet to be understood
if the concept of pseudo-supersymmetry could be extended beyond
cosmologies (time-dependent $(D-2)$-branes) to general
time-dependent $p$-branes (see \cite{Bergshoeff:2007cg} for initial
progress in this direction).

The rest of the paper is organized as follows. In section
\ref{Einstein--Maxwell} we consider Einstein--Maxwell theory and
repeat the construction of the first-order equations for the
non-extremal Reissner-Nordstr\"om black hole. We immediately show
that the same technique allows one to rederive the S0-brane solution
of Einstein--Maxwell theory \cite{Gutperle:2002ai}. In section
\ref{-1-branes} we discuss the special case of $(-1)$-branes in
arbitrary dimensions. In section 4 we explain how the BPS equations
for the $(-1)$-branes imply the BPS equations for general $p$-branes
in arbitrary dimensions via an uplifting procedure. We then discuss
the issue of different black deformations in section \ref{hier} and
finish with conclusions in section \ref{conclusion}.

\section{Four-dimensional Einstein--Maxwell
theory}\label{Einstein--Maxwell}

Einstein-Maxwell theory in four dimensions is described by the action
\begin{equation}\label{Maxwell-Einstein}
S=\int \d^4 x \sqrt{-g} \Bigr(\tfrac{1}{2\kappa^2}\mathcal{R} -
\tfrac{1}{4}F^2\Bigl)\,,
\end{equation}
and has electric and magnetic $0$-branes solutions. Following
\cite{Miller:2006ay} we shall choose a particular Ansatz for the
$0$-brane metric which turns out to be useful
\begin{equation}
\d s^2 = -\epsilon \e^{2A(u)}\d z^2 +
\e^{-2A(u)+2B(u)}\Bigl(\epsilon \e^{2C(u)}\d u^2 +\d \Sigma_k^2
\Bigr)\,.\label{mm}
\end{equation}
If $\epsilon=+1$ then $z$ is time $z=t$ and the metric is
static. For spherical slicings ($k=+1$) this is the appropriate
Ansatz for a black hole, where $u$ is then some function of the
familiar radial coordinate $r$. When $\epsilon=-1$ the metric is
time-dependent and for hyperbolic slicings $(k=-1)$ this is the
appropriate Ansatz for a S$0$-brane \cite{Gutperle:2002ai} with a
one-dimensional Euclidean worldvolume labelled by $z$, and $u$ is
some function of the time-coordinate $\tau$ used in the Milne patch
of Minkowski spacetime. The general Ricci scalar is given by
\begin{equation}
\mathcal{R} = 2\,\epsilon\, \e^{2(A-B-C)}\Bigl(\ddot{A} - \dot{A}^2
+ \dot{A}\dot{B} -\dot{A}\dot{C} -2\ddot{B} + 2\dot{B}\dot{C} -
\dot{B}^2 \Bigr) + 2 k \e^{2(A-B)}\,,
\end{equation}
where a dot indicates a derivative with respect to $u$.

For \emph{electrical} solutions, the Maxwell and Bianchi equations
are solved by
\begin{equation}
F_{uz} = Q \e^{2A-B+C}\,. \label{mfs}
\end{equation}

Plugging the Ans\"{a}tze \rf{mm} and \rf{mfs} into the Einstein
field equations derived from (\ref{Maxwell-Einstein}), one can ask
whether the resulting second-order equations of motion in the one
variable $u$ can be interpreted as field equations for $A, B$ and
$C$ derived from a one-dimensional effective action. It is
straightforward to see that the equations of motion can be obtained
by varying the following action
\begin{equation}\label{blackhole-action}
S=\int \d u\,\,\e^{B-C}\Bigl(2\dot{B}^2-2\dot{A}^2+2\epsilon\,
k\e^{2C} - \epsilon \,\kappa^2 Q^2\e^{2(A-B+C)} \Bigr)\,.
\end{equation}
This action cannot be obtained from direct substitution of the
Ans\"{a}tze in the four-di\-men\-sio\-nal action as the sign of the
resulting $Q^2$-term would be wrong. This sign discrepancy does not
appear for purely magnetic solutions, for which the Ans\"{a}tze can
be plugged into the action consistently. We discuss this point in
detail in appendix A and refer to \cite{VanProeyen:2007pe} for a
careful derivation of the black hole effective action in a more
general setting.

The field $C$ does not appear with a derivative in the action and is
therefore not a propagating degree of freedom. This was to be
expected since $C$ is related to the re-parametrization freedom of
$u$. The field $C$ acts as a Lagrange multiplier enforcing the
following constraint
\begin{equation}\label{constraint}
2\dot{B}^2-2\dot{A}^2-2\epsilon\, k\e^{2C} +
\epsilon \,\kappa^2 Q^2\e^{2(A-B+C)}= 0\,.
\end{equation}
As long as this contraint is satisfied we are free to pick a gauge
choice for $C$ as we like. In the following we choose the gauge
$B=C$.

It turns out that it is easy to generalize the Bogomol'nyi bound
found in \cite{Miller:2006ay} to include both stationary and
time-dependent configurations with arbitrary slicing of the
transverse space $k=0, \pm 1$. The action (\ref{blackhole-action})
is, up to total derivatives, equivalent to
\begin{equation}
S=\int \d u\,\, 2\Bigl(\dot{B}+\sqrt{\epsilon k \e^{2B}
+\beta_1^2}\Bigr)^2 -2\Bigl(\dot{A}+ \sqrt{\epsilon
\tfrac{\kappa^2}{2}Q^2\e^{2A}+\beta_2^2} \Bigr)^2\,, \label{hier2}
\end{equation}
where $\beta_1$ and $\beta_2$ are constants. The BPS equations are
\begin{equation}
\dot{B}=-\sqrt{\epsilon k \e^{2B} +\beta_1^2}\,,\qquad \dot{A}= -
\sqrt{\epsilon \tfrac{\kappa^2}{2}Q^2\e^{2A}+\beta_2^2}\,.
\label{genbps}
\end{equation}
The constraint (\ref{constraint}) implies that
$\beta_1^2=\beta_2^2\equiv \beta^2$. Note that for time-dependent
solutions with charge ($\epsilon=-1$,\quad$Q\neq 0$) the limit of
$\beta_{}\rightarrow 0$ does not exist, while for $Q=0$ the limit
only exists for $k=-1$. The BPS equations are all of the form
\begin{equation}
 \dot D_\pm = - \sqrt{\beta^2 \pm K^2 \e^{2D_\pm}}\,,
\end{equation}
where $K$ is a constant, depending on the case under consideration.
The solutions to these equations are given by
\begin{equation}
\e^{-D_+} = \frac{K}{\beta}\sinh(\beta u + c_+)\,,\qquad\e^{-D_-} =
\frac{K}{\beta}\cosh(\beta u + c_-)\,,
\end{equation}
where $c_\pm$ are constants of integration. In the extremal limit
$\beta \rightarrow 0$ the solution becomes
$\e^{-D_+}=\e^{-D_-}=Ku+c$.

\subsubsection*{Rediscovering Reissner--Nordstr\"om black holes}
For the black hole Ansatz ($\epsilon=+1\,,k=+1$) it was shown in
\cite{Miller:2006ay} that solving the BPS equation described above
leads to the non-extremal Reissner--Nordstr\"om solutions. We shall
now quickly review this for completeness, and draw attention to some
further subtleties.

The solutions of the first-order equations \rf{genbps} are
\begin{equation}
\e^{-A} = \frac{Q\kappa}{\sqrt{2}\beta}\sinh(\beta u +
c_{a})\,,\qquad\e^{-B} = \frac1\beta\sinh \beta u \,,
\label{A}
\end{equation}
where we put the integration constant in the solution for $B$ to
zero by shifting the origin of the $u$-axis, leading to the
following metric
\begin{equation}
\d s^2 = -\frac{2\beta^2}{Q^2\kappa^2\sinh^2(\beta u + c_{a})}\d t^2
+ \frac{\beta^2\kappa^2Q^2\sinh^2(\beta u + c_{a})}{2\sinh^4\beta u}\d u^2
+ \frac{\kappa^2Q^2\sinh^2(\beta u +
c_{a})}{2\sinh^2\beta u }\d \Omega_{2}^2\,.
\end{equation}
We can identify the radial coordinate $r^2$ as the function in front
of $\d\Omega_2^2$. In order to obtain the standard form of the
Reissner-Nordstr\"om solution, one has to perform the following
coordinate transformation:
\begin{equation}\label{transformation}
r= \frac{\kappa Q\sinh(\beta u + c_{a})}{\sqrt{2}\sinh\beta u }\,,
\hspace{2cm} \tau = \frac{\sqrt{2} \beta}{\kappa Q\sinh c_a} t\,,
\end{equation}
such that the solution takes the form
\begin{equation}
\d s^2 = - H(r)\d \tau^2 +  H(r)^{-1}\d r^2 +
r^2\d\Omega_{2}^2\,,\hspace{1cm} F_{\tau r} = -\frac{Q}{r^2},
\label{RN}
\end{equation}
where
\begin{equation}
H(r) = 1 -\frac{2\kappa Q \cosh c_a}{\sqrt{2}r} +
\frac{\kappa^2Q^2}{2 r^2}. \label{H}
\end{equation}
It is then clear that the ADM mass corresponds to $M = {\kappa Q
\cosh c_a}/{\sqrt{2}}$. Note that for $\cosh c_a = 1$, the above
solution reduces to the extreme Reissner-Nordstr\"om metric,
implying that $\cosh c_a$ is related to the non-extremality
parameter $\beta$. Indeed, from (\ref{A}) we have that
\begin{equation}
\cosh c_a = \sqrt{ 1 +  \frac{2 \beta^2 \e^{-2A(0)}}{\kappa^2 Q^2}},
\end{equation}
such that the limit $\cosh c_a = 1$ corresponds to $\beta = 0$, as
one expects from the action (\ref{hier2}).

\subsubsection*{Rediscovering spacelike 0-branes}

For spacelike branes $(\epsilon=k=-1)$ we find
\begin{equation}
\e^{-A} = \frac{\kappa Q}{\sqrt{2}\beta}\cosh(\beta u +
c_{a})\,,\qquad\e^{-B} = \frac1\beta\sinh(\beta u)\,.
\end{equation}
Once again, shifting the origin of the $u$-axis, the integration
constant in the equation for $B$ has been put to zero. Using the
coordinate transformation
\begin{equation}
\tau = \frac{\kappa Q\cosh(\beta u + c_{a})}{\sqrt{2}\sinh\beta u
}\,, \hspace{2cm} x = \frac{\sqrt{2} \beta}{\kappa Q\cosh c_a}
z\label{S0}\,,
\end{equation}
the solution then takes the following form
\begin{equation}
\d s^2 = G(\tau)\, \d x^2 - G(\tau)^{-1}\d \tau^2 + \tau^2\d
\mathbb{H}_2^2\,,\qquad F_{\tau x} = \frac{Q}{\tau^2}\,,
\end{equation}
with
\begin{equation}
G(\tau) = 1 - 2\frac{\sinh(c_a)\kappa Q}{\sqrt{2}\tau}
-\frac{Q^2\kappa^2}{2\tau^2} \,,
\end{equation}
where we introduced the metric for the hyperboloid $\d
\mathbb{H}^2_2 = \d \Sigma^2_{-1}$. Again, this solution is
asymptotically flat. Moreover, we see that this reduces to the
metric for the S0-brane of \cite{Gutperle:2002ai} after a constant
rescaling of $x$ and $\tau$.
Taking the limit $\beta \rightarrow 0$, the metric is easily seen to
describe flat space in Milne coordinates.

\subsubsection*{Addition of a dilaton}

Before we proceed to the case of $p$-branes in arbitrary dimensions
let us first consider the coupling of the vector field to a dilaton, as
this is the generic situation in supergravity theories. The action
describing four-dimensional Einstein-Maxwell-dilaton theory is
\begin{equation}\label{Dilaton-Maxwell-Einstein}
S=\int \d x^4\sqrt{-g} \Bigr(\tfrac{1}{2\kappa^2}\mathcal{R}
-\tfrac{1}{2}(\partial\phi)^2- \tfrac{1}{4}\e^{a\phi}F^2\Bigl)\,.
\end{equation}
The Ansatz for electrical solutions is now given by $F_{uz} = Q
\e^{2A-B+C-a\phi}$. In the gauge $B=C$ the effective action becomes
\begin{equation}
S=\int\d u\,\, 2\dot{B}^2-2\dot{A}^2 - \kappa^2 \dot{\phi}^2 +
2\epsilon\, k\e^{2B} - \epsilon \,\kappa^2 Q^2\e^{2A-a\phi} \,.
\end{equation}
It turns out to be convenient to define new variables $A_1$ and $\phi_1$
\begin{equation}
A_1=A-\frac{a}{2}\phi\,,\qquad \phi_1=\frac a {\kappa^2} A +
\phi\,.
\end{equation}
With these new variables the Bogomol'nyi form is obvious and similar
to the previous case without a dilaton. Writing the action as a sum
of squares, we now introduce three constants $\beta_1,\beta_2$ and
$\beta_3$
\begin{equation}
S = \int \d u\,\,2\Bigl(\dot{B}+\sqrt{\epsilon k \e^{2B}
+\beta^2_1}\Bigr)^2 - \frac2\Delta\Bigl(\dot{A_1}+\sqrt{\epsilon
\Delta\tfrac{\kappa^2}{2}Q^2\e^{2 A_1}+\beta^2_2} \Bigr)^2
-\frac{\kappa^2}{\Delta}(\dot{\phi_1} -
\beta_3)^2,\label{action+dilaton}
\end{equation}
where $\Delta = 1 + (a^2/2\kappa^2)$.

In this case the equivalent of the constraint (\ref{constraint})
implies that only two of the three integration constants are
independent
\begin{equation}
2\beta_1^2 - \frac 2 \Delta\beta_2^2 -\frac{\kappa^2}\Delta
\beta_3^2 =0\,.
\end{equation}
The BPS equations are the same as before apart from the extra
equation $\dot{\phi_1}=\beta_3$.

When the solutions for $A, B$ and $\phi$ are plugged into the Ansatz
one reproduces the familiar dilatonic black hole solution
\cite{Gibbons:1982ih}. One then also notices that the two
independent $\beta$-parameters appear in a fixed combined way as to effectively
form one deformation parameter.

\section{$(-1)$-branes in $D$ dimensions}\label{-1-branes}

A $(-1)$-brane couples electrically to a 0-form gauge potential,
$\chi$, known as the axion. The worldvolume is zero-dimensional and, in
the case of a timelike $(-1)$-brane, this implies that the whole
space is Euclidean since it is entirely transverse. The action is
\begin{equation}
S=\int\d x^D\sqrt{-g}\Big{(} \mathcal{R}-\tfrac{1}{2}(\partial\phi)^2
+\epsilon \tfrac{1}{2}\e^{b\phi}(\partial\chi)^2\Big{)}\,.
\end{equation}
Note the `wrong sign' kinetic term for the axion when $\epsilon=+1$,
which is normal for Euclidean theories. The $(-1)$-brane Ansatz is
\begin{equation}
\d s^2_D=\epsilon \e^{2C(z)} \d z^2 + \e^{2A(z)}\d \Sigma_k^2\,,\hspace{1cm}
\phi= \phi(z)\,,\hspace{1cm}  \chi= \chi(z)\,.
\label{instanton}
\end{equation}
If we consider the axion equation of motion,
$\partial_{\mu}(\sqrt{-g}g^{\mu\nu}\e^{b\phi}\partial_{\nu}\chi)=0$,
then the solution is of the form
\begin{equation}
\dot{\chi}= Q\ \e^{-b\phi}\,.
\end{equation}
The one-dimensional effective action that reproduces the equations
of motion for $A$ and $\phi$ is
\begin{equation}
S=\int \d z \,\tfrac{(D-1)(D-2)}{\kappa^2}\Bigl( \dot{A}^2
\e^{(D-1)A -C} + \epsilon k \e^{(D-3)A +C}\Bigr) - \e^{(D-1)A-C}
\dot{\phi}^2 -\epsilon \e^{C- (D-1)A-b\phi}Q^2\,.
\end{equation}
As we discussed before this form differs from that obtained by
direct substitution of the Ansatz into the original action (appendix
A). The field $C$ is not propagating and we can choose it at will;
the gauge $C=(D-1)A$ is obviously useful. As before we must consider
the constraint that arises from varying the action with respect to
$C$. In this gauge, the BPS form of the action is then equal to
\begin{equation}
S=\int\d z\,\tfrac{(D-1)(D-2)}{\kappa^2}\Bigl( \dot{A}
+\sqrt{\epsilon k \e^{2(D-2)A}+\beta_1^2}\Bigr)^2 - \Bigl(\dot{\phi}
- \sqrt{\epsilon Q^2\e^{-b\phi} +\beta_2^2} \Bigr)^2\,,
\end{equation}
supplemented with the constraint
\begin{equation}
\tfrac{(D-1)(D-2)}{\kappa^2}\Bigl( \dot{A}^2  -\epsilon k
\e^{-2(D-2)A}\Bigr) - \dot{\phi}^2 +\epsilon \e^{-b\phi}Q^2\,= 0.
\end{equation}

The constraint equation tells us that there is only one effective
deformation parameter since
\begin{equation}
\beta_2^2=\tfrac{(D-1)(D-2)}{\kappa^2}\beta_1^2\,.
\end{equation}

We now first solve the BPS equations with vanishing deformation
parameters for $k=\epsilon=1$. If we define the coordinate $\rho$
via $\d \rho=-\e^{(D-1)A}\d z$, then the BPS equation,
$\dot{A}=-\e^{(D-1)A}$, implies that $\rho=\e^A + c$. Shifting
$\rho$ such that $c=0$ we find that the metric describes the
Euclidean plane in spherical coordinates
\begin{equation}
\d s^2_D=\d \rho^2+\rho^2\d \Omega^2_{D-1}\,.
\end{equation}
The solutions for the scalar fields are
\begin{equation}
\e^{\tfrac{b}{2}\,\phi}=-\tfrac{Qb}{2(D-2)}\,\rho^{-D+2} +
\e^{\tfrac{b}{2}\phi_{\infty}}\,,\qquad \chi=-\tfrac{2|Q|}{bQ}
\,(\e^{-\tfrac{b}{2}\phi}  - \e^{-\tfrac{b}{2}\phi_{\infty}})+
\chi_{\infty} \,.
\end{equation}
This is indeed the extremal instanton solution, see for instance
\cite{Gibbons:1995vg,Bergshoeff:2004fq}. For non-zero $\beta$ the
solution becomes (in the frame $C=(D-1)A$)
\begin{align}
& \e^{(2-D)A} = \tfrac{1}{\beta_1}\sinh[(D-2)\beta_1\,z + c_1]
\,,\label{D-1metric}\\
& \e^{-\tfrac{b}{2}\phi(z)}= \tfrac{Q}{\beta_2}\sinh(\tfrac{\beta_2
b}{2}\,z + c_2)\,,\qquad \chi(z) =
-\tfrac{2}{bQ}\sqrt{Q^2\e^{-b\phi}+\beta_2^2} + c_3 \,,
\end{align}
where $c_1, c_2$ and $c_3$ are arbitrary constants of integration.
These solutions correspond to the super-extremal instantons that
were constructed in \cite{Bergshoeff:2004fq, Gutperle:2002km}.

Finally, the time-dependent S$(-1)$ brane solution (with
$k=\epsilon=-1$) that was first constructed in
\cite{Kruczenski:2002ap} can be rederived (again in the frame
$C=(D-1)A$)
\begin{align}
& \e^{(2-D)A} =  \tfrac{1}{\beta_1}\sinh[(D-2)\beta_1\,t + c_1]
\,,\\
& \e^{-\tfrac{b}{2}\phi(t)}= \tfrac{Q}{\beta_2}\cosh(\tfrac{\beta_2
b}{2}\,t + c_2)\,,\qquad \chi(t)= -\tfrac{2}{bQ}\sqrt{Q^2\e^{-b\phi}
- \beta_2^2} + c_3\label{S-1axion}\,.
\end{align}

\section{$p$-branes in arbitrary dimensions}\label{pbranes}
Let us now consider the following theory in $d= D+p+1$ dimensions
\begin{equation}
S=\int \d^{d}x \sqrt{-g}\Bigl\{\tfrac{1}{2\kappa^2}\mathcal{R} -
\tfrac{1}{2}(\partial\phi)^2 -
\tfrac{1}{2(p+2)!}\e^{a\phi}F_{p+2}^2\Bigr\}\,. \label{action2}
\end{equation}

The corresponding $p$-brane solutions can all be reduced to
$(-1)$-brane solutions in $D$ dimensions via reduction over their flat
worldvolumes. Therefore we should be able to reproduce the BPS bounds
and the BPS solutions using the $(-1)$-brane calculation of the
previous section.

A typical $p$-brane Ansatz takes the form
\begin{eqnarray}
&& \d s^2=\e^{2\alpha\varphi(z)}\d s^2_{D} + \e^{2\beta\varphi(z)}\
\eta^\epsilon_{mn}\d y^m\d y^n\,, \hspace{1cm}
 \phi = \phi(z)\,, \nonumber \\
&& A_{p+1}(z)= \chi(z) \ \d y^{1}\wedge\d y^{2}\wedge\ldots\wedge \d
y^{p+1} \,, \label{p-formAnsatz}
\end{eqnarray}
where $\d s^2_D$ is the $D$-dimensional metric (\ref{instanton}) and
$\eta^\epsilon=\text{diag}(-\epsilon, +1,\ldots,+1)$. The constants
$\alpha$ and $\beta$ are given by
\begin{equation}
\label{alpha and beta}
\alpha=\sqrt{\tfrac{p+1}{2(D+p-1)(D-2)}}\,,\qquad
\beta=-\sqrt{
\tfrac{D-2}{2(D+p-1)(p+1)}}\,.
\end{equation}
We now reduce the Ansatz (\ref{p-formAnsatz}) over the transverse coordinates $y$, obtaining a
lower-dimensional Ansatz of the form (\ref{instanton}). The equivalent reduction of the action
(\ref{action2}) leads to the $D$-dimensional action
\begin{equation}
S=\int \d^{D}x \sqrt{-g}\Bigl(\tfrac{1}{2\kappa^2}\mathcal{R} -
\tfrac{1}{2}(\partial\varphi)^2 - \tfrac{1}{2}(\partial\phi)^2 +
\epsilon\tfrac{1}{2}\e^{a\phi+2(D-2)\alpha\varphi}(\partial\chi)^2\Bigr)\,.
\end{equation}
The effective one-dimensional action for the lower-dimensional solution then contains an extra decoupled
dilaton when compared to instanton calculation of the previous section,
\begin{equation}
S=\int \d z\,\tfrac{(D-1)(D-2)}{\kappa^2}\Bigl( \dot{A}^2 + \epsilon
k
 \e^{2(D-2)A}\Bigr)-\dot{\tilde{\phi}}^2-\dot{\tilde{\varphi}}^2 -\epsilon \e^{-b\tilde{\phi}}Q^2\,,
\end{equation}
where $b^2=a^2+4(D-2)^2\alpha^2$ and the original scalars $\varphi$
and $\phi$ are given by
\begin{equation}\label{fieldredef}
\phi=\frac{1}{b}(a\tilde{\phi}-2(D-2)\alpha\tilde{\varphi})\,,\qquad
\varphi=\frac{1}{b}(2(D-2)\alpha\tilde{\phi} + a\tilde{\varphi})\,.
\end{equation}
Up to total derivatives, the BPS-form of the action is then given by
\begin{equation}\label{effective-action2}
S=\int\d z\,\tfrac{(D-1)(D-2)}{\kappa^2}\Bigl(\dot{A}+
\sqrt{\epsilon k g^{2(D-2)}+\beta_1^2}\Bigr)^2 -
\Bigl(\dot{\tilde{\phi}}+ \sqrt{\epsilon
Q\e^{-b\tilde{\phi}}+\beta_2^2} \Bigr)^2 -
(\dot{\tilde{\varphi}}+\beta_3)^2\,,
\end{equation}
where only two of the three deformation parameters $\beta_1$, $\beta_2$ and $\beta_3$
are independent due to the condition coming from the constraint equation:
\begin{equation}
\tfrac{(D-1)(D-2)}{\kappa^2}\beta_1^2 - \beta_2^2 - \beta_3^2=0\,.
\end{equation}
The solutions to the BPS equations for $A$ and $\tilde{\phi}$ can be
found in the previous section in equations
(\ref{D-1metric}-\ref{S-1axion}), whereas the solution for the extra
field $\tilde{\varphi}$ is trivial, $\tilde{\varphi}(z)=-\beta_3 z$.
From our Ansatz (\ref{p-formAnsatz}) and the field redefinition
(\ref{fieldredef}) we can immediately read of the timelike and
spacelike brane solutions in $d$ dimensions. We do not discuss these
solutions as they have been discussed in the literature in numerous
places.

\section{$p$-branes with type II deformations}
\label{hier} In general there are two types of black deformations of
extremal $p$-branes that one can consider \cite{Lu:1996er}. Type I
deformations are defined by the metric
\begin{equation}
\d s^2= \e^{2A(r)}\d \vec{x}^2_{p+1} + \e^{2B(r)}\bigr(\d r^2 +r^2\d
\Omega^2 \bigl)\,.
\end{equation}
The $p$-brane is said to have a type I deformation if
\begin{equation}\label{X}
X \equiv (p+1)A + (D-p-3)B \neq  0\,.
\end{equation}
Since the extremal $(-1)$-brane geometry is the Euclidean plane we
read of that $A=\beta\varphi, B=\alpha\varphi$ such that the
relation between $\alpha$ and $\beta$ (\ref{alpha and beta})
immediately gives $X=0$. From the previous section we can infer that
the cases with $X\neq 0$ can be obtained by uplifting non-extremal
instanton solutions.

However there exist other types of deformations of extremal branes
which are not contained in the analysis of the previous section.
These deformations are labelled type II and the resulting metric
breaks the worldvolume symmetry $\ISO(p,1)$ down to
$\Real\times\ISO(p)$
\begin{equation}
\d s^2= \e^{2A(r)}\bigl(-\e^{2f(r)}\d t^2 +\d x^i\d x^i \bigr) +
\e^{2B(r)} \bigr(\e^{-2f(r)} \d r^2 +r^2\d \Omega^2 \bigl)\,,
\end{equation}
where $X=0$, with $X$ defined in (\ref{X}). For black holes and
instantons these two types of deformations coincide.

The approach of writing the effective action as a sum of squares is
similar to that of the instanton discussions in the previous
section, and is based on dimensionally reducing the brane over its
worldvolume. Notice that although some worldvolume symmetries are
broken we can still carry out the reduction as the translation
symmetry is not broken.

It should be clear that it is possible to reduce type II-deformed
branes over their wordvolume if the shape moduli are not all
truncated. In order to proceed we shall therefore keep a single
shape modulus, denoted by $T$. It appears as follows in the metric
Ansatz
\begin{equation}
\d s^2= \e^{2\alpha\varphi}\d s_D^2 +
\e^{2\beta\varphi}\bigl(-\e^{-T}\d t^2 + \e^{p^{-1}T}\d x_i\d
x^i\bigr)\,.
\end{equation}
The effective action (\ref{effective-action2}) then gets the extra
term $-\tfrac{p+1}{2p}(\dot{T}+\beta_4)^2$ and the corresponding
constraint equation implies the following relation amongst the
deformation parameters
\begin{equation}
\tfrac{(D-1)(D-2)}{\kappa^2}\beta_1^2 - \beta_2^2 - \beta_3^2 -
\tfrac{p+1}{2p}\beta_4^2=0\,.
\end{equation}
Now the various possibilities of choosing non-zero $\beta$'s correspond to
the possible deformations. If all $\beta$'s are non-zero we have a
solution with combined type I and type II deformations. The purely
type II deformed solution can be found by choosing
$\beta_2=\beta_3=0$ with the other two $\beta$'s non-zero. Again
as these solutions can be easily found in the literature we will not
write them explicitly here.

The message here is twofold. Firstly, we have show that the various
types of deformations of $p$-branes can be found from the
first-order formalism. Secondly, this technique is clearly
beneficial in finding various (complicated) black brane solutions,
and is simple in comparison with the existing techniques of solving
the coupled second-order differential equations.

\section{Discussion}\label{conclusion}

In this note we have shown that all known brane-type solutions of an
Einstein-dilaton-p-form theory can be found from decoupled
first-order equations, thereby extending the results of
\cite{Miller:2006ay} to arbitrary dimensions and time-dependent
cases.  With brane-type solutions we mean solutions with a
space-time Ansatz given by (\ref{stationair}) and
(\ref{timedependent}). The key point is that these solutions depend
on one coordinate and can therefore be constructed from a
one-dimensional effective action, as was first discussed for black
holes in \cite{Ferrara:1997tw}. If this one-dimensional effective
action can be written as sums and differences of squares we arrive
at first-order equations \`a la Bogomol'nyi. That this is possible
for some extremal timelike brane solutions was to be expected as
they can be seen as supersymmetric solutions when embedded into an
appropriate supergravity theory.

In \cite{Miller:2006ay} the question was raised as to whether these
deformed BPS equations could also be understood from the point of
view of supersymmetry. One may imagine that the bosonic Lagrangian
(\ref{bosonicaction}) could be embedded into different
(non-standard) supergravity theory for which the non-extremal
solutions preserve some fraction of supersymmetry. However, there is
in fact an obstruction to even defining Killing spinors which
implies that the non-extremal solutions cannot preserve
supersymmetry. Of course one should repeat the same calculations of
\cite{Miller:2006ay} for the case of $p$-branes with $p>0$, as well
as for type I and type II deformations, but we believe that the same
negative answer will be found.

We consider the application of these ideas to time-dependent brane
solutions (S-branes) as a less trivial extension of
\cite{Miller:2006ay}. One possible way to understand why it was to
be expected that a similar first-order formalism exists for
time-dependent branes stems from the known fact that non-extremal
stationary branes can be analytically continued to time-dependent
solutions, something that is impossible for extremal branes
\cite{Lu:1996er}. As explained in the introduction, this first-order
formalism for time-dependent $p$-branes is the natural
generalisation of the so-called pseudo-BPS equations for
FLRW-cosmologies \cite{Skenderis:2006fb, Skenderis:2006jq}.

We did not completely exhaust all possible brane solutions in our
analysis, as we did not consider branes with co-dimensions less than
three . When the co-dimension is one, the stationary branes are
domain walls and the time-dependent branes are FLRW-cosmologies, for
which the fake supergravity and pseudo-supersymmetry formalism is by
now well developed. However, the case of branes with co-dimension
two is not included as these solutions depend on one \emph{complex}
coordinate rather than on one real coordinate.

An alternative, interesting, way to understand the existence of
first-order equations for stationary and time-dependent brane
solutions is given by the approach of mapping $p$-branes to
$(-1)$-branes. The latter solutions are solely carried by the metric
and scalar fields. It is then easy to observe that the scalar fields
only depend on one coordinate and describe a geodesic motion on
moduli space. In fact, for many cases this moduli space is a
symmetric space, for which it is known that the geodesic equation of
motion can easily be integrated to first-order equations (see for
instance \cite{Gaiotto:2007ag}). From this we expect that there
exist BPS equations for all extremal and non-extremal black holes in
theories which have a symmetric moduli space after reduction over
one dimension \cite{future}.

\section*{Acknowledgments}
We would like to thank Jan Perz, Andre Ploegh, Raffaele Punzi and Laura Tamassia for
useful discussions. T.V.R. and B.V. would like to thank the
Departamento de F\'{\i}sica Te\'orica y del Cosmos of the
Universidad de Granada for its hospitality during the last stages of
this work. P.S. would like to thank the K. U. Leuven for its
hospitality during various stages of this work. \\ The work of B.J.
is done as part of the program ``Ram\'on y Cajal'' of the M.E.C.
(Spain). He was also partially supported by the M.E.C. under
contract FIS 2004-06823 and by the Junta de Andaluc\'ia group FQM
101. P.S. is supported by the German Science Foundation (DFG). The
work of T.V.R. and B.V. is supported in part by the FWO -
Vlaanderen, project G.0235.05 and by the Federal Office for
Scientific, Technical and Cultural Affairs through the
Interuniversity Attraction Poles Programme Belgian Science Policy
P5/27. B.V. is also associated to the F.W.O. as Aspirant F.W.O.

\appendix

\section{Construction of the effective action}
As always one has to be careful when plugging an Ansatz into the
action in order to obtain an effective action.

The action contains the following term
\begin{equation}
 S_E = \int \d^4 x f(\psi_i)(F_{uz})^2\,.\label{Elec_Action}
\end{equation}
Here we denoted the other independent fields appearing in the action
by $\psi_i$. For Einstein-Maxwell theory, we have $\{\psi_i\} =
\{g_{\mu\nu}\}$ and $f(g_{\mu\nu}) = -\frac12
\sqrt{-g}g^{uu}g^{zz}$. We continue the discussion keeping $\psi_i$ general. To calculate the contribution of this term to the field
equations for the fields $\psi_i$, the vector field strengths have
to be kept fixed when varying w.r.t. the other fields $\psi_i$.
This gives
\begin{equation}
\frac{\delta S_V}{\delta \psi_i}\Big{|}_{F_{uz}} = \frac{\delta
f}{\delta \psi_i} F_{uz}F_{uz}\label{EOM_SV}
\end{equation}
What happens if we were to plug in the electrical Ansatz? The EOM
are then solved by
\begin{equation}
 F_{uz}(\psi_i) = f^{-1}(\psi_i)Q\,,
\end{equation}
where $Q$ is a constant (the electrical charge). Now the electric
field strength becomes a function of $\psi_i$. This is in contrast
to (\ref{Elec_Action}), where it had to be considered as an
independent field in the action  when calculating the EOM. Due to
this fact, an effective action cannot be obtained by simply
plugging the Ansatz into $S_V$. We have to flip the sign of
$S_V$ too
\begin{equation}
 S_E^{\rm eff} = - S_E (F_{uz}(\psi_i)) = \int \d^4 x (- f^{-1} Q^2)\,.
\end{equation}
The reason is that now we keep $Q$, rather than the field strength
$F_{uz}$, fixed while varying w.r.t. $\psi_i$. In particular, we see
that we gain the correct contribution (\ref{EOM_SV}) to the EOM for
$\psi_i$ by varying the effective action
\begin{equation}
\frac{\delta S_E^{\rm eff}}{\delta \psi_i}\Big{|}_Q = + f^{-2}
\frac{\delta f}{\delta \psi_i} Q^2 = \frac{\delta S_E}{\delta
\psi_i}\Big{|}_{F_{uz}}\,.
\end{equation}

For a magnetic Ansatz we have
\begin{equation}
 F_{z_1 z_2} = P\epsilon_{z_1z_2}\,,\label{Magn_Ansatz}
\end{equation}
where we chose coordinates $z_i$ on the slice $\Sigma_k$. The
contribution to the action for a magnetic configuration is
\begin{equation}
 S_M = \int \d^4x g(\psi_i) (F_{z_1 z_2})^2\,.
\end{equation}
Plugging in this ansatz ({\ref{Magn_Ansatz}}) will not change the EOM
for the fields $\psi_i$, because now the field strength does not
depend on the $\psi_i$.

\bibliography{firstorder}
\bibliographystyle{utphysmodb}

\end{document}